\documentclass[prl,twocolumn,showpacs,preprintnumbers,amsmath,amssymb]{revtex4}
\usepackage{graphicx}
\usepackage{dcolumn}
\usepackage{bm}
\usepackage[tight]{subfigure}
\usepackage{amsmath}
\usepackage{verbatim}
\usepackage{color}

\begin{document}

\title{Quantum information transfer between topological and spin qubit systems}
\author{Martin Leijnse}
\author{Karsten Flensberg}

\affiliation{
  Nano-Science Center \& Niels Bohr Institute,
  University of Copenhagen,
  2100~Copenhagen \O, Denmark 
}

\begin{abstract}
We propose a method to coherently transfer quantum
information, and to create entanglement, between topological qubits and conventional spin qubits.
Our suggestion uses gated control to transfer an electron (spin qubit) between a 
quantum dot and edge Majorana modes in adjacent topological superconductors. 
Because of the spin polarization of the Majorana modes, the electron transfer translates spin superposition states 
into superposition states of the Majorana system, and vice versa.
Furthermore, we show how a topological superconductor 
can be used to facilitate long-distance quantum information transfer and entanglement between spatially separated spin qubits.
\end{abstract}
\pacs{
  03.67.Lx, 
  85.35.Gv, 
  74.45.+c, 
  74.20.Mn, 
}
\maketitle
\emph{Introduction.} 
The main challenge facing the field of quantum computation is the fragile nature of quantum states, i.e., their 
tendency to couple to the environment and decohere into classical states.
In topological quantum computation schemes~\cite{Kitaev03,Nayak08rev}, quantum information is stored in non-local
degrees of freedom which are protected from decoherence stemming from local perturbations. These non-local degrees of freedom can be manipulated by 
braiding operations, i.e., by physical exchange of the associated \emph{local} quasiparticle excitations, if these 
exhibit non-Abelian (non-commutative) statistics~\cite{Stern10rev, Ivanov01}. 
The most prominent candidate for non-Abelian quasiparticles is so-called Majorana modes~\cite{Wilczek09}. These are zero-energy excitations 
existing in vortices or on edges of systems described by a BCS Hamiltonian with $p$-wave type pairing, believed to be realized e.g., 
in the $\nu = 5/2$ fractional quantum Hall state~\cite{Moore91}, in the superconductor Sr$_2$RuO$_4$~\cite{DasSarma06},
and in topological insulators ~\cite{Fu08} or semiconductors with strong spin-orbit 
coupling~\cite{Sau10, Alicea10, Oreg10, Lutchyn10} with induced magnetism and superconductivity. 

However, Majorana modes are examples of Ising anyons~\cite{Bravyi06}, for which braiding operations are not sufficient for universal 
quantum computation~\cite{Nayak08rev}. In general, manipulation and readout of topological qubits appear very challenging
compared to conventional qubit systems. It would therefore be highly desirable to be able to transfer quantum information between
conventional and topological qubits to combine the advantages of each system.
Such interfaces 
have been suggested in the context of anyonic optical 
lattice models~\cite{Jiang08}, and more recently between Majorana modes and superconducting flux qubits~\cite{Hassler10, Sau10b, Jiang11},
and Ref.~\cite{Bonderson11} suggested using an ancillary flux qubit to couple a topological qubit to a qubit 
defined in a double quantum dot.

\begin{figure}[t!]
  	\includegraphics[height=0.2\linewidth]{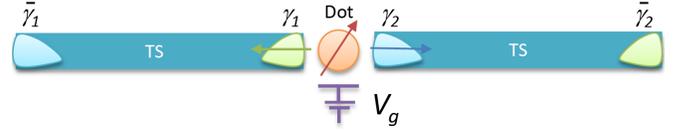}	
	\caption{(Color online) Proposed setup for quantum information transfer between 
	a spin qubit on the quantum dot and a topological qubit, defined within the degenerate ground state 
	of the topological superconductors (TS), with Majorana modes $\gamma_{1,2}$
	on the edges. \label{fig:1}
	The gate voltage $V_g$ controls electron transfer between the dot and TS, thereby transferring
	quantum information between the two systems.
	}
\end{figure}

In this letter, we propose for the first time a
way to directly couple topological and standard spin qubits~\cite{Loss98, Koppens05, Petta05} (in contrast to the suggestion 
in Ref.~\cite{Bonderson11}, we consider qubits based solely on the spin of an electron in a quantum dot, not on the decoherence-prone 
charge component in a double dot).
The suggested setup is sketched in Fig.~\ref{fig:1}, 
where the topological superconductors (TS) are exemplified by wires, but the discussion 
is not limited to this specific case.
A quantum dot is tunnel coupled to two Majorana edge modes, described by the operators 
$\gamma_{1,2}$.
Changing the gate voltage $V_g$ affects a transfer of an electron between the dot and the TS. 
If the Majorana modes are spin-polarized in opposite directions (see below) as indicated by the arrows, the tunneling electron 
must "split", with the spin-up and spin-down components tunneling to/from opposite sides. This coherently transfers a spin superposition
state (spin qubit) into a superposition state within the degenerate ground state described by the Majorana modes, and vice versa.
In this letter, we show in detail how such gate-controlled electron transfer affects qubit transfer between topological 
and spin qubits, and generates entangled spin--topological two-qubit states. In addition, we demonstrate how a multi-dot 
setup can be used to coherently transfer spin qubits between spatially separated quantum dots via the 
TS, and to generate long-distance maximally entangled Bell states between spin qubits.
Quantum dots coupled to Majorana modes were previously considered in Ref.~\cite{Flensberg10b}. 
That work, however, considered spin-polarized dots, which were used only to manipulate the Majorana system, 
not to transfer quantum information between conventional and topological qubits.

\emph{Model.}
Our proposal sets some demands on the experimental realization of the setup sketched in Fig.~\ref{fig:1}.
First, the dot has to be spin degenerate. Since magnetic fields usually cannot be applied locally, the Zeeman term 
needed to induce topological superconductivity could instead be provided by proximity with a magnetic insulator, 
see e.g., Refs.~\cite{Fu08, Sau10}. Alternatively, a magnetic field could be tuned to recover spin degeneracy, but between
spin-up and spin-down states in different dot orbitals.
Second, a Majorana mode is always spin-polarized in the sense that, for a given tunnel junction, it can only 
accept/give off electrons with a specific spin projection: we need this spin polarization to be anti-parallel 
for the $\gamma_1$ and $\gamma_2$ modes.
Under idealized conditions, the spin polarization direction is often controlled 
by the directionality of the border between topological and non-topological regions (and would therefore be opposite for 
opposite edges as sketched in Fig.~\ref{fig:1}),
as discussed e.g., for Majorana bound states in one-dimensional spin-orbit coupled wires~\cite{Oreg10}
or at ferromagnet--superconductor interfaces on the quantum spin Hall edge~\cite{Fu09,Shivamoggi10}, 
and for Majorana edge states, e.g., on the surface of topological insulators 
with induced superconductivity and magnetism~\cite{Fu08}.
Under non-ideal conditions, the polarization direction most likely depends on the details of the edge,
and achieving near perfect anti-parallel spin polarization probably requires
in situ control of some parameter, e.g., the chemical potential near the edges. 
Finally, we need two more Majorana modes in the system, $\bar{\gamma}_1$ and $\bar{\gamma}_2$, 
formed e.g., at the opposite edges of the wires or interfaces, or in vortices in the bulk.

The low-energy Hamiltonian (valid below the superconducting gap $\Delta$) of 
the coupled dot--Majorana system is described by (cf., Refs.~\cite{Tewari08, Flensberg10b, Leijnse11})
\begin{eqnarray}\label{eq:H_QD_MBS}
	H	&=& 	H_D + \gamma_{1} \left( \lambda_1 d_\uparrow - \lambda_1^{*} d_\uparrow^\dagger \right)  + 
			\gamma_{2} \left( \lambda_2 d_\downarrow - \lambda_2^{*} d_\downarrow^\dagger \right),
\end{eqnarray}
where $H_D = \epsilon \sum_\sigma n_\sigma + U n_{\uparrow} n_{\downarrow}$ describes the dot, 
$n_\sigma = d_\sigma^\dagger d_\sigma$ is the number operator for spin projection $\sigma$ (measured along the direction
of the Majorana spin polarization), $\epsilon$ is the on-site energy (controlled 
by $V_g$), and $U$ is the Coulomb charging energy for electrons on the dot.
The two last terms describe tunneling between the dot and Majorana mode $j = 1,2$, with tunnel 
coupling $\lambda_j$, where only spin-up (spin-down) electrons can tunnel into 
mode 1 (2).

A Majorana quasiparticle is its "own hole" and the operators therefore satisfy $\gamma_j^\dagger = \gamma_j$, and 
we assume them to be normalized, $\gamma_j^2 = 1$. It is therefore not possible to count the occupation of a Majorana mode, 
but two Majoranas can be combined to form one ordinary fermion,
$\gamma_{j} = f_{j} + f_{j}^\dagger$, $\bar{\gamma}_{j} = i(f_{j}^\dagger - f_{j})$, 
where $f_{j}^\dagger$ creates a fermion and $f_{j}^\dagger f_j = n_j = 0,1$ counts the occupation of the corresponding state. 
The total system can be described by the number states $| n_1 n_2 D \rangle = | n_1 n_2\rangle_M |D \rangle_D$,
where $D = 0, \uparrow, \downarrow$ describes the dot, which can be empty or occupied by one electron
(we consider only gate voltages where double occupation is suppressed by the large $U$). 

\begin{figure}[t!]
  	\includegraphics[height=0.7\linewidth]{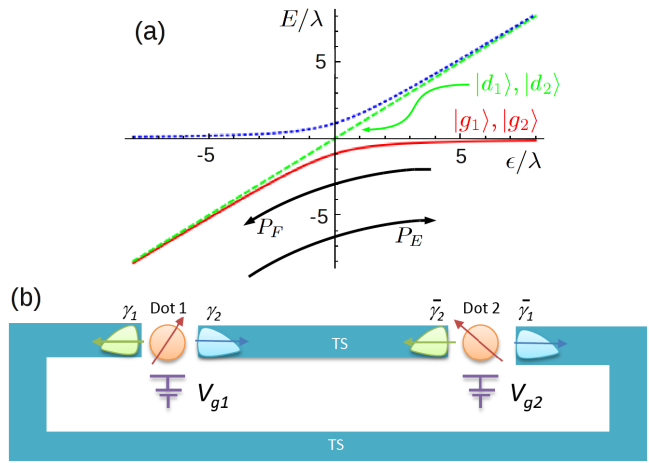}	
	\caption{(Color online) (a) Eigenenergies $E$ of the even parity subspace of Eq.~(\ref{eq:H_QD_MBS}) 
	as function of $\epsilon/\lambda$ in the $U = \infty$ limit. 
	Each eigenstate is two-fold degenerate. $P_E$ and $P_F$ represent operations changing the gate voltage to empty and fill the
	dot, respectively.  
	(b) Sketch of setup with two quantum dots, used for long-distance coherent transfer of spin qubits via 
	the TS and to generate long-distance entanglement of spin qubits.\label{fig:2}
	}
\end{figure}

The Hamiltonian~(\ref{eq:H_QD_MBS}) does not conserve particle number, but the parity (even/odd) of the total number of fermions
(in the dot and the TS) is conserved and we focus for definiteness on the $6$-dimensional even parity subspace. 
Figure~\ref{fig:2}(a) shows the eigenenergies as function of $\tilde{\epsilon} = \epsilon / \lambda$, 
where $\lambda = \sqrt{|\lambda_1|^2 + |\lambda_2|^2}$.
The two degenerate ground states are
\begin{eqnarray}
	\label{eq:groundstates}
	|g_1\rangle	&=&	m(\tilde{\epsilon}) |000\rangle + \frac{\lambda_1^*}{\lambda} 
				n(\tilde{\epsilon}) |10\uparrow\rangle
				+ \frac{\lambda_2^*}{\lambda} n(\tilde{\epsilon}) |01\downarrow \rangle_, \\
	|g_2\rangle	&=&	m(\tilde{\epsilon}) |110\rangle - \frac{\lambda_2^*}{\lambda} 
				n(\tilde{\epsilon}) |10\downarrow\rangle
				+ \frac{\lambda_1^*}{\lambda} n(\tilde{\epsilon}) |01\uparrow \rangle,
\end{eqnarray} 
where $m(\tilde{\epsilon}) = 1 (0)$ when 
$\tilde{\epsilon} \gg 1  (\tilde{\epsilon} \ll -1)$ and $n(\tilde{\epsilon})$ has the opposite behavior,
corresponding to the dot being empty (full) in those limits.
The ground state is degenerate, regardless of the phase difference across the junction,
provided that the spin polarization of the $\gamma_1$ and $\gamma_2$ modes are perfectly anti-parallel,
in which case the ground state energies depend only on $\lambda$.

Now we will exploit this model to transfer quantum information between spin and topological qubits.
The basic operation consists of sweeping $V_g$ to transfer an electron between the dot and the Majorana system. 
In addition, we will need the ability to measure one of the fermion occupation numbers, e.g., $n_1$. How such a measurement is 
done depends on the concrete realization of the TS, see e.g., Refs.~\cite{Kitaev03, Fu08, Akhmerov09, Hassler10}.

\emph{Entangling spin and topological qubits.}
We first assume that the system is prepared in a state $|i_1 \rangle$ with an empty dot  
and a topological qubit defined in the even-parity subspace of the Majorana system ($n_1 + n_2$ even) 
\begin{eqnarray}
	\label{eq:fill1}
	|i_1\rangle = \left( \alpha |00 \rangle_M + \beta |11\rangle_M \right) |0\rangle_D 
		    = \alpha |g_1\rangle + \beta |g_2\rangle,
\end{eqnarray}
where the second equality holds when $\tilde{\epsilon} \gg 1$.
Note that using four Majoranas to define one topological qubit
is standard practice since parity conservation prevents changing the state of a qubit based on only 
two Majoranas~\cite{Bravyi06}.
Next, we increase the gate voltage so as to move down the dot level to 
$\tilde{\epsilon} \ll -1$. If this gate sweep is done adiabatically, the system remains in 
the same superposition of the ground states, which is now however (up to an overall dynamical phase factor)
\begin{eqnarray}
	\label{eq:fill2}
	|i_1\rangle 	&\rightarrow&	\frac{\lambda_1^*}{\lambda} \left( \alpha |10\rangle_M + 
					\beta|01\rangle_M \right) |\uparrow\rangle_D \nonumber \\  
			&+&		\frac{\lambda_2^*}{\lambda} \left( -\beta |10\rangle_M 
					+ \alpha|01\rangle_M \right) |\downarrow\rangle_D.					
\end{eqnarray}
This is already a very interesting result: 
by simply sweeping the gate voltage so as to fill the dot with an electron, the spin-dependence of the tunneling 
terms have given rise to an \emph{entangled two-qubit state}, involving the quantum dot spin and a topological 
qubit defined within the odd-parity sector of the Majorana system. Note that if the initial state is either $|g_1\rangle$ or
$|g_2\rangle$, and if $|\lambda_1| = |\lambda_2|$, the result is a maximally entangled 
Bell state.
We denote by $P_F$ the operation of filling the dot with an electron by an adiabatic gate sweep.
Up to an overall phase, this has the same effect as acting on a state with an empty dot with the operator 
\begin{eqnarray}
	\label{eq:fill}
	P_F	&=&	\frac{1}{\lambda} \left( -\lambda_1^* \gamma_{1} d^\dagger_\uparrow - 
			 \lambda_2^* \gamma_{2} d^\dagger_\downarrow \right). 
\end{eqnarray}

\emph{Transferring topological to spin qubits.}
Starting from the entangled two-qubit state in Eq.~(\ref{eq:fill2}) we can measure e.g., $n_1$, collapsing 
the Majorana wavefunction to either
$N |01\rangle_M (\lambda_1^* \beta |\uparrow\rangle_D + \lambda_2^*\alpha|\downarrow\rangle_D)$ if the result is $n_1 = 0$,
or $N |10\rangle_M (\lambda_1^* \alpha |\uparrow\rangle_D - \lambda_2^* \beta|\downarrow\rangle_D)$, if $n_1 = 1$, 
where $N$ is a normalization factor. 
These states describe a spin qubit on the dot, which, if $|\lambda_1| = |\lambda_2|$, is related to the 
initial topological qubit state by a unitary transformation. Knowing the result of the $n_1$ measurement and 
the relative phase of $\lambda_1$ and $\lambda_2$, we know the transformation and have thus coherently transferred 
the topological qubit we started with in $|i_1\rangle$ to a spin qubit.

\emph{Transferring spin to topological qubits.}
Next we consider the opposite operation: We start with the Majorana system prepared in a known state, say $n_1 = 1, n_2 = 0$, and 
an occupied dot, where the spin qubit is in some arbitrary state,
$|i_2\rangle = |1 0 \rangle_M (a |\uparrow \rangle_D + b | \downarrow \rangle_D )$.  
For $\tilde{\epsilon} \ll -1$ this initial state written in the eigenbasis is
\begin{eqnarray}
	\label{eq:save1}
	|i_2\rangle &=&  \frac{1}{\lambda} \left( a \lambda_1 |g_1 \rangle - b \lambda_2 |g_2 \rangle
			- a \lambda_2^* |d_1 \rangle + b \lambda_1^* |d_2 \rangle \right), 
\end{eqnarray}
where 
$|d_1\rangle =  ( -\lambda_2 |10 \uparrow \rangle + \lambda_1 |0 1 \downarrow \rangle )/\lambda$ and 
$|d_2\rangle =  ( \lambda_1 |10 \downarrow \rangle + \lambda_2 |0 1 \uparrow \rangle )/\lambda$
are the eigenstates with energy equal to $\epsilon$ in Fig.~\ref{fig:2}(a), which 
have no component with an empty dot. Thus, even if $V_g$ is swept adiabatically to $\tilde{\epsilon} \gg 1$, the dot may
not be empty. This is in fact to be expected since the dimension of the Hilbert space with a filled dot is twice 
as large as that with an empty dot.  
We assume that the charge on the dot is measured after the gate sweep, and initially assume that this measurement indicates
an empty dot, and show how to deal with the case of a filled dot below. 
After the charge measurement, the state is therefore given by the $|g_1\rangle$ and $|g_2\rangle$ components of 
Eq.~(\ref{eq:save1}) only. We denote by $P_E$ the operation of
adiabatically sweeping the gate from $\tilde{\epsilon} \ll -1$ to $\tilde{\epsilon} \gg 1$, followed by a charge measurement which is 
assumed to show an empty dot. 
Up to a phase, this is equivalent to acting on a state with a filled dot with 
\begin{eqnarray}
	\label{eq:empty}
	P_E 	&=&	N \left( \lambda_1 \gamma_{1} d_\uparrow + \lambda_2 \gamma_{2} d_\downarrow \right), 
\end{eqnarray}
where the normalization factor $N$ depends on the initial state because of the 
dot charge measurement, associated with a partial collapse of the wavefunction. The effect on the state $| i_2 \rangle$ is
\begin{eqnarray}
	\label{eq:save2}
	P_E |i_2 \rangle &=&	N \left( -a \lambda_1 |  0 0 \rangle_M + b \lambda_2 |1 1 \rangle_M \right) |0\rangle_D.
\end{eqnarray}
If $|\lambda_1| = |\lambda_2|$, this describes a unitary (coherent) transfer of the spin qubit to a topological 
qubit.

If the charge measurement following the gate sweep instead indicates a filled dot, the resulting state  
is given by the $|d_1\rangle$ and $|d_2\rangle$ components of Eq.~(\ref{eq:save1}):
\begin{eqnarray}
	\label{eq:decoupled_state}
	|i_2 \rangle	&\rightarrow&	N \left[ \left( a |\lambda_2|^2 |10\rangle_M + 
					b \lambda_1^* \lambda_2 |0 1 \rangle_M \right) |\uparrow\rangle_D \right. \nonumber \\
			&+&	 	\left. \left( b |\lambda_1|^2 |10\rangle_M 
					-a \lambda_2^* \lambda_1 |0 1 \rangle_M \right) |\downarrow\rangle_D \right],
\end{eqnarray}
which is an entangled two-qubit state. 
Transferring the spin qubit to the Majorana system can still be achieved, 
most straightforwardly by re-initializing the whole system and trying again.
Alternatively, we can measure e.g., $n_1$, whereupon the state in Eq.~(\ref{eq:decoupled_state}) collapses to 
$N |10\rangle_M ( a |\lambda_2|^2  |\uparrow\rangle_D + b  |\lambda_1|^2 |\downarrow\rangle_D )$ if $n_1 = 1$  and 
$N |01\rangle_M ( b \lambda_1^* \lambda_2 |\uparrow\rangle_D - a \lambda_2^* \lambda_1 |\downarrow\rangle_D)$ if $n_1 = 0$.
Thus, we can recover the initial spin qubit and try the transfer operation again. 
The last option is to measure the dot spin, causing a collapse to either 
the first or the second line in Eq.~(\ref{eq:decoupled_state}), which
has already affected a transfer of the spin qubit to a topological qubit, but defined 
within the odd-parity subspace of the Majorana system (leaving the dot occupied, but with a known spin projection).

\emph{Experimental considerations.} Importantly, the adiabatic charge-transfer operations discussed above
are insensitive to environmental degrees of freedom coupling to the dot charge~\cite{Flensberg10b}.
To avoid involving quasiparticle excitations of the TS or
doubly occupied dot states, we need $\lambda \ll \Delta , U$. In most quantum dot realizations $U$ can be several meV. 
The effective $\Delta$ in the topological phase depends on the system at hand, 
but assuming $\Delta \sim 0.1-1$~meV would make $\lambda \sim 1-10~\mu$eV sufficient. 
Adiabatic operation requires changing
$\epsilon$ on a time scale slower than $1/\lambda$ ($\sim 0.1-1$~ns with the above estimate, 
which is fast compared to typical spin qubit dephasing times~\cite{Petta05, Koppens05}). 
In fact, adiabatic gate-sweeps on such time scales are commonly used in spin qubit readout~\cite{Petta05} 
(spin-to-charge conversion).
In addition, there are experimental setups where it is possible to control the tunnel 
couplings via electrostatic gates~\cite{Nadj10} and $\lambda$ could then be changed 
together with $\epsilon$ to more easily achieve 
$|\tilde{\epsilon}| \gg 1$ at the end and beginning of operations.
A finite spin-splitting $\delta B$ on the dot does not break the two-fold degeneracy of the eigenstates 
of~(\ref{eq:H_QD_MBS}) since the Majorana system still provides two 
degenerate states. However, for $\tilde{\epsilon} \ll -1$, $\delta B$ splits off the two-fold 
degenerate ground state $|g_{1,2}\rangle$ from the states $|d_{1,2}\rangle$, which now have the character of
spin-up/-down along the field direction, and adiabatically filling the dot will always result in the same 
state of the spin qubit. A gate sweep which is non-adiabatic with respect to $\delta B$ (but adiabatic 
with respect to $\lambda$) still creates the above entangled state, and therefore small splittings,
$\delta B \ll \lambda$, are acceptable.
If the spin polarizations of the $\gamma_1$ and $\gamma_2$ modes are not perfectly anti-parallel, the fidelity 
of the qubit transfer is reduced. In addition, the degeneracy of the ground state is lifted, unless the 
relative phase across the junction is tuned to exactly $\pi$~\cite{Flensberg10b}, 
$\text{arg} \; \lambda_1 - \text{arg} \; \lambda_2 = \pi$.

\emph{Transferring spin qubits.}
We now consider a modified 
setup, involving a second quantum dot, see Fig.~\ref{fig:2}(b), and
use number states $| n_1 n_2\rangle_M |D_1 D_2 \rangle_D$ with obvious notation. It is a simple generalization of the above 
arguments to define operations $P_{F2}$ and $P_{E2}$, adiabatically filling and emptying the second dot. 
Consider now the initial state 
$|i_3\rangle = |1 0 \rangle_M (a |\uparrow 0\rangle_D + b | \downarrow 0\rangle_D )$,
i.e., a spin qubit in the first dot and 
an empty second dot. Emptying the first dot [see Eq.~(\ref{eq:save2})] and then filling the second one gives
\begin{eqnarray}
	\label{eq:extractD2}
	P_{F2} P_{E1} |i_3 \rangle &=& 
		N \left[ \left( a \lambda_1 \bar{\lambda}_2^* | 0 1 \rangle_M - 
		b \lambda_2 \bar{\lambda}_2^* |1 0 \rangle_M \right) |0 \uparrow \rangle_D \right. \nonumber \\
		&+& \left. \left(  b \lambda_2 \bar{\lambda}_1^* | 0 1 \rangle_M +
		a \lambda_1 \bar{\lambda}_1^* |1 0 \rangle_M \right) 
		|0 \downarrow \rangle_D \right], 
\end{eqnarray}
where $\bar{\lambda}_{1,2}$ are the tunnel couplings between the Majorana modes $\bar{\gamma}_{1,2}$ and dot 2.
The spin on the second dot has now been entangled with the Majorana system and measuring e.g., $n_2$
affects a transfer of the spin qubit from the first to the second dot.
In principle this transfer can be long-ranged, but it is limited by the ability to initialize and measure the non-local
Majorana pairs $\gamma_1,\bar{\gamma}_1$ and $\gamma_2,\bar{\gamma}_2$. This could for example be done in a network 
of one-dimensional superconducting wires, where the Majorana bound states can be moved around with
gates~\cite{Alicea10b}.

\emph{Generating long-distance entanglement.} 
As a final example, we start from $|i_4 \rangle = |0 0\rangle_M |0 0\rangle_D$, i.e., both dots are empty, and fill first 
dot 1 and then dot 2 
\begin{eqnarray}
	\label{eq:extractD1D2_1}
	P_{F2} P_{F1} |i_4 \rangle &=&  
			N \left[ |1 1 \rangle_M \left( \lambda_1^* \bar{\lambda}_2^* |\uparrow \uparrow \rangle_D 
				- \lambda_2^* \bar{\lambda}_1^* |\downarrow \downarrow \rangle_D \right) \right. \nonumber \\ 
		&+&   	\left. |0 0 \rangle_M \left( \lambda_1^* \bar{\lambda}_1^* |\uparrow \downarrow \rangle_D + 
				\lambda_2^* \bar{\lambda}_2^* |\downarrow \uparrow \rangle_D \right) \right].  
\end{eqnarray}
Simply by subsequently filling the two dots we have generated an entangled 
three-qubit state. If we now measure e.g., $n_2$
we generate two spatially separated entangled spin qubits. Note that if each dot has equal
tunnel couplings to both Majorana modes, $|\lambda_1| = |\lambda_2|$ and $|\bar{\lambda}_1| = |\bar{\lambda}_2|$,
the spin qubits always end up in a Bell state.
Thus, TS can be used to affect quantum teleportation of electron spins. 

\emph{Conclusions.} We have presented a method to 
transfer quantum information between, and to entangle, topological qubits and conventional spin qubits.
This allows using well-established experimental techniques for quantum computations with spin qubits, while 
topological qubits are used as quantum memories for long-time storage. 
Alternatively, our proposal can be used to implement partially protected \emph{universal topological quantum computation}.
In fact, braiding operations can implement a universal set of one- and two-qubit gates, 
provided that certain ancillary one- and two-qubit states can be prepared~\cite{Bravyi06}. The required two-qubit state can be prepared in a 
topologically protected way~\cite{Sau10b}, but the single-qubit state needed is more problematic.
In our setup, this (and any other topological single-qubit state) can be prepared with high accuracy by initializing the dot spin 
in the appropriate angle relative to the Majorana spin polarization and then emptying the dot, see Eq.~(\ref{eq:save2}).

In addition, we have shown that a topological superconductor can be used to transfer spin qubits between
spatially separated quantum dots, and to create long-distance entanglement between spin qubits. 
Thus, topological superconductors can act as buses for quantum information in conventional qubit systems, 
or facilitate quantum teleportation of electron spins.

We thank A. Reynoso for discussions.

\bibliographystyle{apsrev}
\end{document}